\documentstyle[12pt]{article}
\title{One variation on Lloyd's theme}
\author{
     Oleg Derzhko\\
{\em Institute for Condensed Matter Physics}\\
{\em 1 Svientsitskii St., L'viv-11, 290011, Ukraine}\\
[5pt]and\\
[5pt]
     Johannes Richter\\
{\em Institut f\"{u}r Theoretische Physik,
     Universit\"{a}t Magdeburg}\\
{\em P.O.Box 4120, D-39016, Magdeburg, Germany}}
\date{\today}

\begin{document}

\maketitle

\begin{abstract}
\normalsize
One random spin-$\frac{1}{2}$ $XY$ chain
that after Jordan-Wigner fermionization reduces to the extended Lloyd's
model is considered. The random-averaged one-fermion Green functions
have been calculated exactly
that yields thermodynamics of the spin model.

\end{abstract}

\vspace{2mm}

\noindent
{\em PACS codes:} 75.10.-b

\vspace{2mm}

\noindent
{\em Keywords:}
Spin-$\frac{1}{2}$ $XY$ chain;
Lorentzian disorder;
Green functions approach;
Density of states;
Thermodynamics;
Magnetization;
Susceptibility\\

\vspace{2mm}
\noindent
{\bf Postal addresses:}\\

\vspace{1mm}
\noindent
{\em
Dr. Oleg Derzhko (corresponding author)\\
Institute for Condensed Matter Physics\\
1 Svientsitskii St., L'viv-11, 290011, Ukraine\\
Tel: (0322) 427439\\
Fax: (0322) 761978\\
E-mail: derzhko@icmp.lviv.ua\\

\vspace{1mm}
\noindent
Prof. Dr. Johannes Richter\\
Institut f\"{u}r Theoretische Physik, Universit\"{a}t Magdeburg\\
P.O.Box 4120, D-39016, Magdeburg, Germany\\
Tel: 0049 391 671 2473\\
Fax: 0049 391 671 1131\\
E-mail: johannes.richter@physik.uni-magdeburg.d400.de}

\clearpage
\renewcommand\baselinestretch 2
\large\normalsize

An idea to exploit Lloyd's model \cite{l}
for examining the thermodynamical properties of random
spin-$\frac{1}{2}$ $XY$ chains belongs to
H.Nishimori \cite{ni}.
He noted that after Jordan-Wigner trick \cite{lsm}
the Hamiltonian of
isotropic $XY$ model with random lorentzian transverse field
describes
tight-binding spinless fermions
with diagonal
lorentzian disorder. Since the
random-averaged
one-fermion Green functions for
such model were found exactly by P.Lloyd,
one can obtain the thermodynamics of
random spin system via the averaged density of states.
Later the treatment presented in \cite{ni}
was generalized for the cases of alternating bonds
\cite{o} and
additional intersite Dzyaloshinskii-Moriya interaction \cite{dv}.

On the other hand,
W.John and J.Schreiber suggested an
extension of Lloyd's method to
off-diagonal disorder \cite {js} that was successfully used in the
study of disordered systems
%\cite{rhs,rsh,r1,r2,hk}.
[7-11].
The idea of the present communication is to exploit Lloyd's
model with off-diagonal disorder
for analysis of thermodynamics of
the corresponding random spin-$\frac{1}{2}$ $XY$ chain.
Similarly to \cite{ni} we were able to calculate exactly various
thermodynamical quantities, although found somewhat different results of
influence of randomness on these functions.

We consider $N$ spins
$\frac{1}{2}$ arranged in a circle with the Hamiltonian
\begin{eqnarray}
H= \sum_{n=1}^N \Omega_n s_n^z
+\sum_{n=1}^N J_n \left( s^x_ns^x_{n+1}+s^y_ns^y_{n+1}\right) ,
\;\;\;s_{n+N}^{\alpha}=s_n^{\alpha},
\end{eqnarray}
where $\Omega_n$ is a transverse field at site $n$ and $J_n$ is
the interaction between the sites $n$ and $n+1$.
The latter are taken to be random with a probability distribution density
\begin{eqnarray}
p(J_1,...,J_N)=
\prod _{n=1}^N\frac{1}{\pi}\frac{\Gamma}
{\left( J_n-J_0\right)^2+\Gamma^2},
\end{eqnarray}
that is the product of lorentzian distribution densities at sites that
are centered at $J_0$ with the width $\Gamma$.
In order to treat the model (1), (2) in exact manner
the transverse field $\Omega_n$ at each site must depend on surrounding
intersite interactions in the following way
\begin{eqnarray}
\Omega_n-\Omega_0=
a\left( \frac{J_{n-1}-J_0}{2}+ \frac{J_{n}-J_0}{2}\right),
\;\;\;
a \; {\mbox {is real,}}
\;
\mid a \mid \ge 1,
\end{eqnarray}
where $\Omega_0$ is the averaged transverse field at site.

Really, by Jordan-Wigner transformation
from operators
$s^{\pm}_j\equiv s^x_j\pm is^y_j$
to Fermi operators
$c_j$,
$c_j^+$
the Hamiltonian (1) becomes
\begin{eqnarray}
H=H^-+BP^+,
\nonumber\\
H^{-}\equiv -\frac{1}{2}\sum_{n=1}^N \Omega_n
+\sum_{n=1}^N \Omega_n c^+_nc_n
+\sum_{n=1}^N \frac{J_n}{2} \left( c^+_nc_{n+1}-c_nc^+_{n+1}\right) ,
\;\;\;c_{n+N}=c_n, \;\;c_{n+N}^+=c_n^+,
\nonumber\\
B\equiv -J_N\left( c^+_Nc_{1}-c_Nc^+_{1}\right) ,
\;\;\;P^{+}\equiv \frac{1+P}{2}, \;\;\; P\equiv
\prod_{n=1}^N\left( -2s_n^z\right) .
\end{eqnarray}
For calculation of thermodynamical properties of the model (1)
one can omit the boundary term $B$ \cite{sm},
and hence
one faces with one-dimensional version of Anderson's model with
the off-diagonal disorder considered by W.John and J.Schreiber.

In order to study thermodynamics one should
diagonalize the bilinear in Fermi operators form $H^-$ (4)
by canonical transformation
$\eta_k=\sum_{n=1}^Ng_{kn}c_n$
with real $g_{kn}$ that satisfy the equations
$\Lambda_kg_{kn}=\sum_{i=1}^Ng_{ki}A_{in}$
with
$A_{ij}\equiv \Omega_i\delta_{ij}+
\frac{1}{2}J_i\delta_{j,i+1}+
\frac{1}{2}J_{i-1}\delta_{j,i-1}$,
and
$\sum_{i=1}^Ng_{ki}g_{pi}=\delta_{kp}$,
$\sum_{p=1}^Ng_{pi}g_{pj}=\delta_{ij}$
obtaining in result
$H^-=\sum_{p=1}^N \Lambda_p (\eta^+_p\eta_p - \frac{1}{2})$.
The density of states
$\rho (E) \equiv \frac{1}{N} \sum_{p=1}^N \delta (E-\Lambda_p)$
determines thermodynamics for certain realization of random intersite
interactions. For example, the Helmholtz free energy per site is given by
$f=-\frac{1}{\beta}\int dE \rho (E)
\ln (2{\mbox {ch}}
\frac{\beta E}{2})$.
The Helmholtz free energy averaged over random realizations is
given by the same formula only with the
random-averaged density of states
$\overline{\rho (E)}$,
where the averaging is defined by
$\overline{(...)}\equiv \int dJ_1...dJ_Np(J_1,...,J_N)(...)$.

On the other hand, the temperature double-time Green functions
$\Gamma_{pq}^{\mp}(t) \equiv
\mp i \theta (\pm t)<\{ \eta_p(t),\eta^+_{q}\} >$,
$\Gamma^{\mp}_{pq}(t)=\frac{1}{2\pi}
\int_{-\infty}^{\infty}dE {\mbox {e}}^{-iEt}
\Gamma^{\mp}_{pq}(E \pm i \varepsilon )$,
$\varepsilon \rightarrow +0$
yield the density of states for a certain random realization:
$\rho (E)=\frac{1}{N}\sum_{p=1}^N \left[
\mp \frac{1}{\pi}{\mbox {Im}}\Gamma^{\mp}_{pp}(E\pm i\varepsilon )\right]$.
$\rho (E)$
can be rewritten in terms of Green
functions
$G^{\mp}_{nm}(t) \equiv \mp i \theta (\pm t)<\{ c_n(t),c^+_{m}\} >$
as
$\rho (E)=
\mp \frac{1}{\pi}
\frac{1}{N}\sum_{j=1}^N
{\mbox {Im}}G^{\mp}_{jj}(E\pm i\varepsilon )$
since
$\Gamma_{pq}^{\mp}(t)=
\sum_{i=1}^N\sum_{j=1}^Ng_{pi}g_{qj}G^{\mp}_{ij}(t)$.
In result the averaged density of states is determined by the averaged
Green functions $\overline{G^{\mp}_{nm}(E)}$ via the relation
$\overline{\rho (E)} =
\mp \frac{1}{\pi }{\mbox {Im}}\overline{G^{\mp}_{nn}(E)}$.

Finally, following \cite{js} one can derive the exact expression
for $\overline{G^{\mp}_{nm}(E)}$.
First it is necessary to write a set of equations for
$G^{\mp}_{nm}(E\pm i \varepsilon )$
that follows from equations of motion for
$G^{\mp}_{nm}(t)$ and
then to average these
equations using contour integration in complex
planes of $J_n$s. Under the imposed condition (3)
on the basis of Gershgorin criterion one can state that
for $a\ge 1$
the retarded (advanced) Green function
cannot have a pole in lower (upper) half-planes of $J_n$s,
whereas for $a\le -1$
in upper (lower) half-planes of $J_n$.
Therefore,
every contour of integration should be closed in the half-plane
where there is only the pole originated from lorentzian distribution, and
after trivial use of residues one gets a set of equations for the averaged
Green functions that possess already the translational symmetry and
hence may be solved in a strandard way. The final result for the
averaged Green functions reads
\begin{eqnarray}
\overline{G^{\mp}_{nm}(E)}=
\frac
{\left( \frac{\sqrt{x^2-y^2}-x}{y}\right)^{\mid n-m\mid}}
{\sqrt{x^2-y^2}}
\end{eqnarray}
with
$x \equiv E - \Omega_0 \pm i\mid a \mid \Gamma$,
$y \equiv J_0 \mp i {\mbox {sgn}}(a)\Gamma $.

The obtained averaged Green functions (5) permit to study thermodynamics of
spin model (1)-(3). Really, the required averaged density of states that
follows from (5) reads
\begin{eqnarray}
\overline{\rho (E)} =
\mp \frac{1}{\pi }{\mbox {Im}}
\frac{1}{ \sqrt{ \left( E-
\Omega_0 \pm i\mid a \mid \Gamma\right)^2-
\left(J_0 \mp i{\mbox {sgn}}(a)\Gamma\right)^2}}
\nonumber\\
=\frac{1}{\pi}
\sqrt{\frac{\sqrt{A^2+B^2}-A}
{2(A^2+B^2)}},
\nonumber\\
A\equiv (E-\Omega_0)^2+(1-\mid a \mid ^2)\Gamma^2-J_0^2,
\;\;\;
B\equiv 2\Gamma [\mid a \mid (E-\Omega_0)+{\mbox {sgn}}(a)J_0].
\end{eqnarray}
The entropy and specific heat can be calculated by formulae
\begin{eqnarray}
\overline{s}=
\int dE \overline{\rho (E)}\left[ \ln{
\left(2{\mbox{ch}}\frac{\beta E}{2}\right)}
-\frac{\beta E}{2}{\mbox{th}}\frac{\beta E}{2}\right],
\\
\overline{c}=
\int dE \overline{\rho (E)}
\left(
\frac{\frac{\beta E}{2}}
{{\mbox {ch}}{\frac{\beta E}{2}}}\right)^2.
\end{eqnarray}
Due to the noteworthy property of (6)
$\frac{\partial}{\partial \Omega_0}\overline{\rho (E)}=
-\frac{\partial}{\partial E}\overline{\rho (E)}$
one can express transverse magnetization and static transverse
linear susceptibility through the density of states
\begin{eqnarray}
\overline{m_z}\equiv
\overline{<\frac{1}{N}\sum_{n=1}^Ns_n^z>}=
-\frac{1}{2}\int dE \overline{\rho (E)}{\mbox{ th}}\frac{\beta E}{2},
\\
\overline{\chi_{zz}}\equiv
\frac{\partial \overline{m_z}}{\partial \Omega_0}=
-\beta \int dE \overline{\rho (E)}\frac{1}
{(2{\mbox {ch}}{\frac{\beta E}{2}})^2}.
\end{eqnarray}

Let us discuss the obtained results.
In the absence of randomness ($\Gamma =0$)
(6) reduces to the well-known result:
$\overline{\rho (E)}=
\frac{1}{\pi} \frac{1}{\sqrt{J_0^2-(E-\Omega_0)^2}}$
if
$\mid E-\Omega_0\mid \le \mid J_0\mid$
and
$\overline{\rho (E)}=0$ otherwise.
The isotropic $XY$ model in random lorentzian
transverse field treated by
H.Nishimori may be obtained in the limit
$\Gamma \rightarrow 0$,
$\mid a \mid \Gamma ={\mbox {const}}=\Gamma_{{\mbox {N}}}$.
The model in question (1)-(3) essentially differs from
that model:
the density of states (6)
in contrast to the case of diagonal disorder is
not symmetric with respect to the change
$E-\Omega_0 \rightarrow - (E-\Omega_0)$.
However, it remains the same after the replacement
$E-\Omega_0 \rightarrow - (E-\Omega_0)$,
$a \rightarrow -a$,
or
$E-\Omega_0 \rightarrow - (E-\Omega_0)$,
$J_0 \rightarrow -J_0$,
since the simultaneous change of signs of $J_0$ and $a$
in (6) does not affect $\overline{\rho (E)}$.
For convenience hereafter
will be put $J_0=1$.
The above-mentioned symmetry
of the density of states
can be seen in Fig.1,
where the averaged density of states (6) for $\Gamma =1$
is displayed. The
density of states for non-random case
is depicted in Fig.1 by dashed lines.
For large $\mid a\mid$ due to disorder
the edges of the zone are completely smeared out;
for $\mid a\mid \approx 1$
the disorder results in smearing out mainly of one edge of
the zone.
Some consequences induced by this dependence of
$\overline{\rho (E)}$ on $a$ for $\Gamma \neq 0$
will be seen in the behaviour of thermodynamical quantities.

The results of numerical calculations
of thermodynamical quantities
for $\Gamma =1$ and few values of $a$
are presented in Figs.2-5,
namely,
the temperature dependences of entropy (7) (Fig.2),
specific heat (8) (Fig.3) and
static transverse linear susceptibility (10) (Fig.5) and
the dependence on averaged transverse field at low temperatures
of the transverse magnetization (9) (Fig.4);
the curves that correspond to non-random case
are depicted in these figures by dashed lines.
The influence of randomness on thermodynamics is mainly rather typical.
It leads to weak deformation of the curve entropy
versus temperature
with decreasing of entropy at high temperatures (Fig.2),
broadening and decreasing
of the peak in dependence specific heat versus
temperature (Fig.3),
smearing out of the cast in the $\overline{m_z}$ versus
$\Omega_0$ curve at $T=0$ for $\Omega_0=J_0$
and nonsaturated transverse magnetization at any finite transverse field
(Fig.4),
suppressing of static transverse linear susceptibility
versus temperature curve (Fig.5).
However, as can be seen in Figs.2-5 the
influence of disorder, especially for small $a$,
essentially depends on the sign of $a$.
Particularly interesting is the case of strong asymmetry
in the density of states
$\overline{\rho (E)}$
when $\mid a \mid \approx 1$.
From mathematical point of view the dependence of computed quantities on
temperature and averaged transverse field and the well-pronounced
difference
between the cases $a\approx -1$ and $a\approx 1$
can be understood
while bear in mind that these quantities according to (7)-(10)
are the integrals over $E$ of the products of $\overline{\rho (E)}$
depicted in Fig.1 by the functions with evident dependence on $E$ at
different $\beta$.
It is interesting to note that for some Hamiltonian parameters and
temperatures even the large randomness (controlled by $\Gamma$) almost
does not affect the observable thermodynamical quantities.
This can be nicely seen in Figs.2-5.

It is worth to underline that the asymmetry of
$\overline{\rho (E)}$
leads to the appearance of nonzero transverse magnetization
$\overline{m_z}$ at zero averaged transverse field $\Omega_0$.
As it can be seen from (9)
$\overline{m_z}=0$
at $T=0$, $\Omega_0=0$
if
$\int_{-\infty}^0dE\overline{\rho (E)}=
\int_0^{\infty}dE\overline{\rho (E)}$.
This is evidently true for a symmetric density of states
$\overline{\rho (E)}$
(as in the case considered by H.Nishimori)
but is not obvious in the case in question (6).
The difference between the integrals
$\int_{-\infty}^0dE\overline{\rho (E)}$
and
$\int_0^{\infty}dE\overline{\rho (E)}$
can be clearly demonstrated by numerical finite-chain calculations
\cite{dk}
as a difference between the numbers of negative and positive
eigenvalues of $N\times N$ matrix $\mid \mid A_{ij}\mid \mid$
$\Lambda_p$,
denoted by
${\cal {N}}_-$
and
${\cal {N}}_+$
respectively,
for certain realization of random model (1)-(3).
For a realization of random chain (1)-(3) of 1000 spins with
$\Omega_0=0$, $J_0=1$, $\Gamma =1$ that gives
$\frac{1}{N}\sum_{n=1}^NJ_n=1.009$
we found that
for $a=-5$
${\cal {N}}_-=495$,
${\cal {N}}_+=505$,
for $a=-2$
${\cal {N}}_-=470$,
${\cal {N}}_+=530$,
for $a=-1.01$
${\cal {N}}_-=402$,
${\cal {N}}_+=598$.
Another random realization of this chain with
$\frac{1}{N}\sum_{n=1}^NJ_n=0.986$
yields
for $a=-5$
${\cal {N}}_-=503$,
${\cal {N}}_+=497$,
for $a=-2$
${\cal {N}}_-=471$,
${\cal {N}}_+=529$,
for $a=-1.01$
${\cal {N}}_-=408$,
${\cal {N}}_+=592$.
The transverse magnetization for certain realization at $T=0$ is given by
$\overline{m_z}=\frac{{\cal {N}}_--{\cal {N}}_+}{2N}$
and one finds a good agreement of calculated in such a manner
$-\overline{m_z}$
with the results depicted in Fig.4.

To summarize,
this paper is devoted to thermodynamics of spin-$\frac{1}{2}$
isotropic $XY$ chain with random lorentzian intersite interaction and
transverse field that depends linearly on the surrounding intersite
interactions (1)-(3).
The derived exact expressions
for the averaged density of states (6) and thermodynamical
quantities (7)-(10)
seems to be interesting from academic point of view
since they permit to understand the disorder effects and from applied
point of view since they may be used as
a testing ground for approximate
methods of spin systems with off-diagonal disorder.

Unfortunately, the obtained results do not permit to calculate exactly
the averaged spin correlation functions because such calculation
requires the
knowledge of averaged many-particle fermion Green functions.
Spin correlations and their dynamics may be examined
using exact finite-chain calculations developed in
\cite{dk2,dkv}.

\vspace{5mm}

One of the authors (O.D.)
would like to thank to T.Krokhmalskii and T.Verkholyak
for helpful discussions.
He is grateful to
the Deutscher Akademischer Austauschdienst for
a scholarship for stay in
Germany when the present study was started.
He is also indebted to 
Mr. Joseph Kocowsky for
continuous financial support.
The work was partly supported
by the Deutsche Forschungsgemeinschaft (Project
Ri 615/1-2).

\clearpage
\input{F1.PIC}

\clearpage
\input{F2.PIC}

\clearpage
\input{F3.PIC}

\clearpage
\input{F4.PIC}

\clearpage
\input{F5.PIC}

\clearpage

\noindent
{\bf List of figure captions}\\
\vspace{0.25cm}

\noindent
Fig.1.
The averaged density of states (6)
$\overline{\rho (E)}$ vs. $E-\Omega_0$.
\vspace{0.5cm}

\noindent
Fig.2.
The entropy $\overline{s}$ (7) vs. temperature $\frac{1}{\beta}$.
\vspace{0.5cm}

\noindent
Fig.3.
The specific heat $\overline{c}$ (8) vs. temperature $\frac{1}{\beta}$.
\vspace{0.5cm}

\noindent
Fig.4.
The transverse magnetization $-\overline{m_z}$ (9) vs.
transverse field $\Omega_0$
at low temperature ($\frac{1}{\beta}=0.001$).
\vspace{0.5cm}

\noindent
Fig.5.
The static transverse linear susceptibility
$-\overline{\chi_{zz}}$ (10) vs. temperature
$\frac{1}{\beta}$ at $\Omega_0=0.5$.
\vspace{0.5cm}


\begin{thebibliography}{99}
\bibitem{l} P.Lloyd, J.Phys.C 2 (1969) 1717.
\bibitem{ni} H.Nishimori, Phys.Lett.A 100 (1984) 239.
\bibitem{lsm} E.Lieb, T.Schultz, D.Mattis, Ann.Phys. 16 (1961) 407.
\bibitem{o} K.Okamoto, J.Phys.Soc.Jap. 59 (1990) 4286.
\bibitem{dv} O.Derzhko, T.Verkholyak, internal report of
ICTP IC/95/182 (Miramare-Trieste, 1995).
\bibitem{js} W.John, J.Schreiber, phys.stat.sol.(b) 66 (1974) 193.
\bibitem{rhs} J.Richter, K.Handrich, J.Schreiber,
phys.stat.sol.(b) 68 (1975) K61.
\bibitem{rsh} J.Richter, J.Schreiber, K.Handrich,
phys.stat.sol.(b) 74 (1976) K125.
\bibitem{r1} J.Richter, phys.stat.sol.(b) 87 (1978) K89.
\bibitem{r2} J.Richter, phys.stat.sol.(b) 99 (1980) K13.
\bibitem{hk} K.Handrich, S.Kobe,
Amorphe Ferro- und Ferrimagnetika
(Akademie-Verlag, Berlin, 1980)
(in German).
\bibitem{sm} Th.J.Siskens, P.Mazur, Physica A 71 (1974) 560.
\bibitem{dk} O.Derzhko, T.Krokhmalskii, Ferroelectrics 153 (1994) 55.
\bibitem{dk2} O.Derzhko, T.Krokhmalskii, JMMM 140-144 (1995) 1623.
\bibitem{dkv} O.Derzhko, T.Krokhmalskii, T.Verkholyak, JMMM 157-158 (1996) .
\end{thebibliography}
\end{document}